\documentclass[letterpaper]{article} 
\usepackage{aaai2026}  
\usepackage{times}  
\usepackage{helvet}  
\usepackage{courier}  
\usepackage[hyphens]{url}  
\usepackage{graphicx} 
\usepackage{natbib}  
\usepackage{caption} 
\usepackage{algorithm}

\usepackage{newfloat}
\usepackage{listings}
\lstdefinelanguage{json}{
  basicstyle=\footnotesize\ttfamily,
  showstringspaces=false,
  morestring=[b]",
  morecomment=[l]{//},
  moredelim=[l]{:},
  stringstyle=\ttfamily
}
\DeclareCaptionStyle{ruled}{labelfont=normalfont,labelsep=colon,strut=off} 
\floatstyle{ruled}
\newfloat{listing}{tb}{lst}{}
\floatname{listing}{Listing}
\DeclareUnicodeCharacter{FF0C}{,}  
\DeclareUnicodeCharacter{FF08}{(}  
\DeclareUnicodeCharacter{FF09}{)}  
\usepackage{amsfonts}
\usepackage{booktabs}
\usepackage{tabularx, array}
\usepackage{enumitem}
\usepackage[noend]{algpseudocode}
\usepackage{amsmath} 

\usepackage{tikz}
\usetikzlibrary{shapes.geometric, arrows.meta, positioning, calc, fit}

\usepackage{makecell}          

\newcommand{\rowdivider}{\cmidrule{1-2}}  
\newcolumntype{L}{>{\raggedright\arraybackslash}p{0.36\linewidth}}
\newcolumntype{R}{>{\raggedright\arraybackslash}X}

\usepackage{xcolor}

\title{Scaling Equitable Reflection Assessment in Education via Large Language Models and Role-Based Feedback Agents}
\author{
    Chenyu Zhang\textsuperscript{\rm 1}\thanks{Corresponding author.},
    Xiaohang Luo\textsuperscript{\rm 2}
}

\affiliations{
    \textsuperscript{\rm 1}Harvard University\\
    \textsuperscript{\rm 2}University of Pennsylvania\\
    chenyu\_zhang@alumni.harvard.edu, xiaohl@upenn.edu
}

\begin{document}

\maketitle

\begin{abstract}
Formative feedback is widely recognized as one of the most effective drivers of student learning, yet it remains difficult to implement equitably at scale. In large or low-resource courses, instructors often lack the time, staffing, and bandwidth required to review and respond to every student reflection, creating gaps in support precisely where learners would benefit most.
This paper presents a theory-grounded system that uses five coordinated role-based \textsc{LLM} agents (Evaluator, Equity Monitor, Metacognitive Coach, Aggregator, and Reflexion Reviewer) to score learner reflections with a shared rubric and to generate short, bias-aware, learner-facing comments.
The agents first produce structured rubric scores, then check for potentially biased or exclusionary language, add metacognitive prompts that invite students to think about their own thinking, and finally compose a concise feedback message of at most 120 words. The system includes simple fairness checks that compare scoring error across lower and higher scoring learners, enabling instructors to monitor and bound disparities in accuracy.
We evaluate the pipeline in a 12-session AI literacy program with adult learners. In this setting, the system produces rubric scores that approach expert-level agreement, and trained graders rate the AI-generated comments as helpful, empathetic, and well aligned with instructional goals.
Taken together, these results show that multi-agent \textsc{LLM} systems can deliver equitable, high-quality formative feedback at a scale and speed that would be impossible for human graders alone. 
The approach demonstrates how structured agent roles, fairness checks, and learning-science principles can work together to support instructors while preserving pedagogical intent. More broadly, the work points toward a future where feedback-rich learning becomes feasible for any course size or context, advancing long-standing goals of equity, access, and instructional capacity in education.
\end{abstract}

\begin{links}
    \link{Code, Prompts, and Anonymized Data}{https://github.com/CharlieChenyuZhang/equitable-reflection-assessment}
\end{links}

\section{Introduction}
Timely, high-quality formative feedback\,\citep{juwah2004enhancing,pishchukhina2021supporting,morris2021formative} is one of the most powerful levers for
closing achievement gaps in education, yet it remains out of reach for large
classes and low-resource programs.
Decades of scholarship show that rich, dialogic feedback can raise learning gains by up to 0.7 effect-size points
\citep{black1998inside,hattie2007power}.  Unfortunately, instructors cannot
read and respond to every learner reflection when enrolments scale into the
hundreds or when teaching assistants are scarce. The resulting feedback gap disproportionately undermines the academic growth and motivation of historically marginalized students, further exacerbating existing disparities in educational attainment \citep{nicolai2023literature}.

Large language models (LLMs) offer a tantalising alternative: they can read and
comment on text at super-human speed without task-specific fine-tuning. Early studies already suggest that GPT-4 can achieve human-level agreement in automated essay scoring when supplied with appropriate rubrics and prompt constraints \citep{garcia2025chatgpt}. 
However, when an explicit, teacher-designed rubric is absent, LLMs have been
shown to over-emphasize superficial presentation cues or implicitly apply
value-laden constructs that fall outside the course’s intended learning
outcomes. 
These risks highlight the necessity of grounding any AI-assisted grading workflow in a shared reference framework, whether through detailed rubrics, constrained prompting, or post-hoc human calibration, to ensure that automated evaluations reinforce rather than undermine pedagogical objectives
\citep{ouyang2023systematic,brown2022past}.

We address these gaps with a theory-grounded, multi-agent pipeline that (1) produces \emph{robust, equitable} rubric scores and (2) delivers \emph{dialogic} feedback aligned with learning-science theory.
By integrating advances in natural language processing with principles from the learning sciences, our approach provides formative feedback that is accurate and scalable while remaining pedagogically grounded and attentive to equity.

\section{Background \& Related Work}

\begin{quote}
\emph{“Learning without reflection is a waste. Reflection without learning is dangerous.”}\\[2pt]
Confucius
\end{quote}

\paragraph{Formative Feedback}
Formative assessment serves as an ongoing gauge of student learning, supplying timely feedback rather than waiting for end-of-course summative judgments. High-quality formative feedback promotes self-regulated learning and narrows
achievement gaps when it is specific, actionable, and dialogic
\citep{juwah2004enhancing,pishchukhina2021supporting,morris2021formative}.
Many studies have examined the use of formative feedback during the writing process. Providing students with formative feedback while they write is a key instructional practice that helps them improve as writers \citep{graham2011informing,macarthur2016instruction}. By clearly communicating what high-quality performance looks like and how to achieve it, formative feedback directs students toward productive action and improvement in specific writing skills \citep{graham2012meta,panadero2023effects}.
To alleviate the time‐intensive nature of essay evaluation, researchers have applied natural language processing and artificial intelligence techniques \citep{grimes2010utility,graham2015formative,roscoe2017presentation}. However, these systems remain less reliable than expert human raters, and developing them demands extensive technical expertise and large, prompt‐specific training corpora \citep{chen2022examining,moore2016student}.
Recent work has started exploring the use of GPT-4 to generate written feedback for English language learners, highlighting its potential to provide timely, detailed, and structured responses \cite{carlson2023utilizing}.

\paragraph{Automated Grading Systems.}
Automated grading has a long history in educational technology, originating in the 1960s with formative work on programming assignment evaluation \citep{hollingsworth1960automatic} and automatic essay scoring \citep{page1966imminence}.
Early systems relied on deterministic pipelines, including unit-testing harnesses for code \citep{messer2024automated}, rule-based heuristics \citep{liu2019automatic,ureel2019automated}, and ensemble strategies that used stacking with domain adaptation to transfer across tasks \citep{heilman2013ets}.
The rapid progress of deep learning has fundamentally reshaped this landscape: neural models now routinely outperform traditional pipelines in both reliability and generalization \citep{riordan2017investigating}. 

The rapid expansion of large language models (LLMs) in natural language processing has sparked their integration into automated grading pipelines, where prompt-based evaluation is aligned with explicit marking schemes. Researchers have successfully applied few-shot fine-tuning or in-context demonstration to a range of state-of-the-art models, including BERT \citep{sung2019pre}, and GPT-4 \citep{chiang2024large,impey2025using,xiao2025human}.
Across a range of essay and short-answer benchmarks, prior work has demonstrated grading accuracy that is comparable to, and in some cases better than, traditional baselines, with expert educators validating many of these gains. In comparison, fully \emph{zero-shot} prompting is still under-examined and has not yet shown the same level of reliability.
Early investigations using simple prompts on university-level courses show promise for low-stakes formative tasks yet fall short on comprehensive summative assessments such as final examinations \citep{kortemeyer2023can,yeung2025zero}.

\paragraph{Gap and Contribution.}
Most LLM-based grading systems concentrate on overall score agreement and leave broader social-impact goals unaddressed. These goals include equitable treatment of learners at different proficiency levels, feedback that supports motivation and self-regulation, and workflows instructors can audit, interpret, and adapt. Prior work on multi-agent or equity-aware LLM pipelines has examined cultural alignment, feedback quality, or bias detection in isolation, rather than as part of an end-to-end reflection-assessment system suitable for real classrooms. To our knowledge, no existing approach integrates stable rubric scoring, bias-aware dialogic feedback, and explicit fairness evaluation for reflective writing into a single, scalable workflow.
\textbf{This work contributes:}
(i) a scalable, self-consistent scoring pipeline that produces auditable rubric scores with minimal human supervision; and
(ii) a role-based agent ensemble that generates bias-aware, conversational feedback aligned with formative-assessment and metacognitive principles.

\section{Problem Formulation}
\label{sec:problem}
High-quality, \emph{dialogic} formative feedback is one of the most effective ways to narrow achievement gaps, yet instructors in large or low-resource courses cannot realistically review and respond to every learner reflection. Addressing this challenge requires an approach that produces rubric-aligned scores that are accurate and fair, along with narrative comments that are pedagogically useful for learners, all without increasing the workload placed on instructors or creating new barriers for students who are already underserved. To formalize this challenge, we describe the reflection-assessment task and introduce notation that supports precise evaluation of scoring accuracy, equity, and feedback quality.

\paragraph{Task Definition.}
We cast reflection assessment as a \textbf{two-output prediction task} with
explicit equity constraints:

\begin{enumerate}[label=(\alph*), leftmargin=1.6em,itemsep=2pt]
  \item a \emph{rubric scorer}
        $f:\mathbb{X}\!\to\![0,3]$ that predicts an ordinal score for each of
        the four dimensions in Table~\ref{tab:rubric-grad}; and
  \item a \emph{feedback generator}
        $g:(x_i,f(x_i))\!\to\!\mathbb{T}$ that emits a concise,
        learner-facing comment $t_i\in\mathbb{T}$ of no more than
        120 words, aligned with the five quality criteria in
        Table~\ref{tab:rubric-fb}.
\end{enumerate}

\paragraph{Notation.}
Let $\mathcal{D} = \{(x_i, y_i)\}_{i=1}^{N}$ be a corpus of $N$ reflections where:
\begin{itemize}[leftmargin=1.2em,itemsep=1pt]
  \item $x_i \in \mathbb{X}$ represents the reflection text, and
  \item $y_i \in \{0,\dots,3\}^{4}$ represents expert rubric scores on four dimensions.
\end{itemize}

A disjoint evaluation set
$\mathcal{E}=\{(x_j,y_j,q_j)\}_{j=1}^{M}$ additionally contains
$q_j\in[1,5]^5$, the mean Likert ratings (Correctness, Alignment, Actionability, Depth, Empathy) that three trained human graders assign to the
AI-generated comment for reflection $x_j$.

\paragraph{Objective 1: Scoring Accuracy.}
We quantify how closely the system’s predictions align with expert labels
using two complementary statistics: absolute deviation and ordinal agreement.

\begin{enumerate}[leftmargin=1.6em,itemsep=2pt]

    \item \textbf{Mean Absolute Error (MAE).}
          Let the rubric comprise \(D = 4\) scored dimensions.
          For reflection \(x_i\), denote the model-predicted score on
          dimension \(d\) by \(f_{d}(x_i)\) and the corresponding
          majority-vote human score by \(y_{i,d}\).
          The per-dimension MAE is
    
          \begin{align}
            \operatorname{MAE}_{d} &=
              \frac{1}{N}\sum_{i=1}^{N}
              \bigl|f_{d}(x_i) - y_{i,d}\bigr|,
              \label{eq:mae-dim}\\[4pt]
            \operatorname{MAE}_{\text{overall}} &=
              \frac{1}{D}\sum_{d=1}^{D}\operatorname{MAE}_{d},
              \label{eq:mae-overall}
          \end{align}
    
          where Eq.\,\eqref{eq:mae-dim} averages the absolute error for
          a \emph{single} rubric dimension and
          Eq.\,\eqref{eq:mae-overall} averages those \(D\) values to
          yield one scalar.  Lower scores indicate tighter
          model–human alignment; \(\operatorname{MAE}=0\) denotes perfect
          scoring fidelity.

  \item \textbf{Quadratic weighted kappa (QWK).}
        Let \(C=\{0,\dots,K\}\) be the set of ordinal score categories,
        \(O_{ij}\) the observed co-occurrence matrix,
        \(E_{ij}\) the expected matrix under independence, and
        \(w_{ij}=\dfrac{(i-j)^2}{(K-1)^2}\) the quadratic weight.  
        The kappa statistic is
        \begin{equation}
          \label{eq:qwk}
          \operatorname{QWK}
          = 1 - \frac{\displaystyle\sum_{i,j} w_{ij}\,O_{ij}}
                       {\displaystyle\sum_{i,j} w_{ij}\,E_{ij}} ,
        \end{equation}
        ranging from \(-1\) (complete disagreement) to \(1\) (perfect
        agreement); higher values signify stronger ordinal concordance.

\end{enumerate}

\paragraph{Objective 2: Equity.}
Even though we do not collect sensitive demographic attributes, we can still
operationalize \emph{fair treatment} by requiring the scorer to be equally
accurate for learners at different proficiency levels.
Concretely, we follow established grading-fairness research that compares performance across \emph{score bands} \citep{liu2016fairness}.
Let $\mathcal{B}$ partition $\mathcal{D}$ into two bands based on the
\emph{human} rubric score $y_i$:  
\textit{low} ability (0–1 points) and \textit{high} ability (2–3 points).
Fairness is quantified by the worst-band error gap
\begin{equation}
  \label{eq:delta_mae}
  \Delta_{\text{MAE}}
  = \max_{b\in\mathcal{B}}
      \bigl|\operatorname{MAE}_{b}(f)
            - \operatorname{MAE}_{\neg b}(f)\bigr| .
\end{equation}
where $\operatorname{MAE}_{b}$ is computed over reflections whose true score
falls in band $b$. Minimizing $\Delta_{\text{MAE}}$ ensures that lower-scoring learners, who are most in need of formative support, are not systematically under- or over-scored by the automated system.

\paragraph{Objective 3: Feedback Usefulness.}
For each generated comment
$t_j = g\,\bigl(x_j,f(x_j)\bigr)$,
three trained graders rate its quality on the five
Likert-scaled dimensions in Table~\ref{tab:rubric-fb},
yielding a vector $q_j \in [1,5]^5$.
Define the aggregate quality score
\begin{equation}
  \label{eq:q_quality}
  Q(g) \;=\;
  \frac{1}{M}\sum_{j=1}^{M}
  \frac{1}{5}\sum_{d=1}^{5} q_{j,d}\, .
\end{equation}

where $M = |\mathcal{E}|$ and $d$ indexes the rubric dimensions
\textit{(Correctness, Alignment, Actionability, Depth, Empathy)}.
The objective is to \textbf{maximize} $Q(g)$ while
(i) honoring the 120-word length cap and
(ii) ensuring that neither the overall error
$\operatorname{MAE}(f)$ nor the equity gap
$\Delta_{\text{MAE}}$ degrades as defined in Objectives 1–2.

\paragraph{Research Questions.}
Guided by the goals of accuracy, equity, and practicality, we investigate three core research questions:
\begin{enumerate}[label=\textbf{RQ\arabic*:},labelwidth=2.3em,labelsep=0.5em,leftmargin=2.8em,itemsep=2pt]
  \item \textbf{Scoring Accuracy and Fairness}  
        Can a zero-shot LLM scorer match expert graders on both the overall \(\operatorname{MAE}\) and the equity gap \(\Delta_{\text{MAE}}\)?
  \item \textbf{Feedback Usefulness}  
        Does the role-based agent ensemble achieve a mean quality rating of \(Q(g)\!\ge\!4.0\) on the five-point scale?
 \item \textbf{Practicality}  
        Are the time and cost requirements of the pipeline low enough to support deployment in large or resource-constrained courses?
\end{enumerate}

\noindent
No reinforcement learning or fine-tuning is performed.  
All automation relies on \emph{prompt-based} GPT-4o calls whose outputs are evaluated against \emph{human-expert rubrics} for both scoring and feedback quality, aligning the evaluation with the social-impact emphasis on equity, transparency, and pedagogical value.

\section{Methodology}
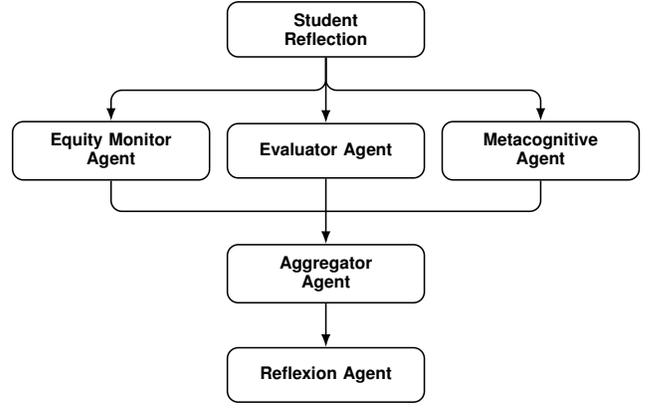
\begin{figure}[t]
    \centering
    \resizebox{\linewidth}{!}{%
    \begin{tikzpicture}[
        block/.style={
            rectangle, 
            draw=black, 
            thick, 
            rounded corners=2mm, 
            text width=3.2cm, 
            minimum height=1.0cm, 
            align=center, 
            font=\sffamily\bfseries\footnotesize, 
            inner sep=6pt
        },
        line/.style={
            draw, 
            thick, 
            -Latex, 
            rounded corners=2mm
        },
        connector/.style={
            draw, 
            thick,
            rounded corners=2mm
        }
    ]

    
    \node [block] (reflection) {Student\\Reflection};

    \node [block, below=1.2cm of reflection] (evaluator) {Evaluator Agent};

    \node [block, left=0.3cm of evaluator] (equity) {Equity Monitor\\Agent};
    \node [block, right=0.3cm of evaluator] (metacog) {Metacognitive\\Agent};

    \node [block, below=1.2cm of evaluator] (aggregator) {Aggregator\\Agent};

    \node [block, below=0.8cm of aggregator] (reflexion) {Reflexion Agent};


    \draw [line] (reflection.south) -- (evaluator.north);
    \coordinate (splitpoint) at ($(reflection.south)!0.5!(evaluator.north)$);
    \draw [line] (reflection.south) -- (splitpoint) -| (equity.north);
    \draw [line] (reflection.south) -- (splitpoint) -| (metacog.north);

    \draw [line] (evaluator.south) -- (aggregator.north);
    \coordinate (mergepoint) at ($(evaluator.south)!0.5!(aggregator.north)$);
    \draw [connector] (equity.south) |- (mergepoint);
    \draw [connector] (metacog.south) |- (mergepoint);

    \draw [line] (aggregator.south) -- (reflexion.north);

    \end{tikzpicture}
    } 
    \caption{Role-based, multi-agent workflow for equitable reflection assessment.}
    \label{fig:pipeline}
\end{figure}

Our pipeline is implemented as a role-based, multi-agent
workflow orchestrated with AutoGen \citep{wu2024autogen} (Figure~\ref{fig:pipeline}). Five
specialized GPT-4o agents collaborate asynchronously so that each reflection
emerges with both per-dimension rubric scores and a concise, dialogic feedback
comment. The execution logic is summarized in Algorithm~\ref{alg:feedback}.

\subsection{Role Configuration}
Each agent role is motivated by principles from formative assessment, equity-centered pedagogy, and multi-agent LLM design. The workflow separates responsibilities to promote transparency, reduce compounding errors, and allow instructors to audit or adjust individual components. The five roles are:

\begin{enumerate}[label=\arabic*.]

\item \textbf{Evaluator Agent} applies the four dimension rubric to the raw reflection text and returns a structured JSON object. The output includes (i) an integer score from 0 to 3 for each rubric dimension, (ii) a brief natural language explanation of the score in a \texttt{reasoning} field, and (iii) an \texttt{areas\_for\_improvement} list that highlights specific suggestions for the learner. This structured format provides an explicit, auditable record of how each score is assigned and why.

\item \textbf{Equity Monitor Agent} reviews the Evaluator’s narrative for biased, exclusionary, or culturally insensitive phrasing and proposes revisions. This step supports equitable treatment across learners by reducing unintended linguistic or evaluative bias.

\item \textbf{Metacognitive Agent} generates one or two reflective prompts that encourage the learner to examine their reasoning and plan next steps. This aligns the workflow with research in self-regulated learning and dialogic feedback.

\item \textbf{Aggregator Agent} synthesizes the outputs of the previous agents into a concise, learner-facing comment capped at 120 words. By highlighting only a small number of actionable next steps, the Aggregator helps prevent feedback overload \citep{underwood2006improving,black2009developing,grimes2010utility}.

\item \textbf{Reflexion Agent} performs a lightweight post-hoc check, returning \textsc{CONFIDENT} or \textsc{REVISE} along with targeted suggestions. This final layer improves reliability by catching inconsistencies or omissions before feedback is released.

\end{enumerate}

Generation temperature is fixed at \(0.3\) to balance determinism and nuance.

\vspace{4pt}
\begin{algorithm}[t]
\footnotesize            
\caption{Four-Agent Feedback Pipeline}
\label{alg:feedback}
\begin{algorithmic}[1]
  \Require reflection $x$, rubric $R$
  \Ensure  score $s$, comment $t$
  \State $\mathcal{A}\gets\{\text{Evaluator},\text{Equity},\text{Meta}\}$
  \ForAll{$a\in\mathcal{A}$ \textbf{in parallel}}
      \State\hspace{1em}$o_a\gets a.\Call{Generate}{x,R}$
  \EndFor
  \State $s\gets\Call{ParseScore}{o_\text{Evaluator}}$
  \State $t\gets\Call{Aggregator.Combine}{%
        o_\text{Evaluator},\,o_\text{Equity},\,o_\text{Meta}}$
  \State $(c,\delta)\gets\Call{Reflexion}{t}$
  \If{$c=\text{REVISE}$}
      \State $t\gets\Call{Revise}{t,\delta}$
  \EndIf
  \State \Return $(s,t)$
\end{algorithmic}
\end{algorithm}

\section{Experiments}

This section details the empirical set-up used to validate the proposed
reflection-assessment pipeline.

\paragraph{Dataset.}
\label{subsec:dataset}
Our analysis draws on 336 written reflections produced by 28 adult learners enrolled in an online, synchronous AI literacy program (approximately 12 reflections per learner). Participants ranged in age from 18 to 22 years. All reflections were written in American English, collected under an approved IRB protocol, and released with informed participant consent.

\paragraph{Evaluation Metrics.}
\label{subsec:metrics}
We evaluate scoring accuracy, fairness, and feedback quality using four complementary families of metrics:

\begin{enumerate}[label=(\arabic*), leftmargin=1.6em, itemsep=4pt]

    \item \textbf{Score alignment.}
          Agreement between the \emph{system} score vector and the \emph{human-reference} vector (the majority vote across three rubric-trained annotators) is assessed using two statistics:
          \begin{itemize}[leftmargin=1.4em,itemsep=1pt]
            \item \textbf{Mean Absolute Error} (\textbf{MAE}), defined in Eqs.~\eqref{eq:mae-dim} and~\eqref{eq:mae-overall}, which measures the average $L_{1}$ deviation across the four rubric dimensions.
            \item \textbf{Quadratic Weighted Kappa} (\textbf{QWK}), defined in Eq.~\eqref{eq:qwk}, which quantifies chance-corrected ordinal concordance.
          \end{itemize}

    \item \textbf{Inter-rater reliability.}
          We compute the two-way mixed intraclass correlation coefficient $\mathrm{ICC}^{(2,1)}$ for both Human–Human and Human–AI pairs to assess consistency at the item level.

    \item \textbf{Fairness disparity.}
          Following Eq.~\eqref{eq:delta_mae}, we measure the error gap $\Delta_{\text{MAE}}$ across learner proficiency bands to evaluate whether scoring accuracy is comparable for lower- and higher-scoring learners.

    \item \textbf{Feedback quality.}
          We summarize grader judgments using the aggregate usefulness score \(Q(g)\) defined in Eq.~\eqref{eq:q_quality}, which averages ratings across the five feedback-quality dimensions.
\end{enumerate}

\paragraph{Participants and Compensation.}
Three human annotators were recruited and compensated for their work. Each annotator received a unique Google Sheet along with an instructional video that explained the annotation procedure, rubric dimensions, and task expectations. Annotators were required to view the full video before beginning the task.

\paragraph{Sampling and Annotation Design.}
To balance temporal coverage with cognitive load, we sampled reflections from three key points in the course: Class 1 (early), Class 6 (midpoint), and Class 12 (final). For each of the 28 learners, one reflection from each selected class was included, yielding a total of \(3 \times 28 = 84\) reflections. Each annotator completed six tasks: grading human-written reflections from Classes 1, 6, and 12, and evaluating AI-generated feedback for the same three classes. All annotations were submitted directly in the assigned Google Sheet using the shared four-dimension rubric.

\paragraph{Rubrics.}
The four-dimension grading rubric (Table~\ref{tab:rubric-grad}) was identical
for humans and AI.  After viewing each AI-generated comment, annotators rated
its quality on five Likert-scaled dimensions
(Table~\ref{tab:rubric-fb}); anchors followed best practice in
educational-feedback research.

\begin{table}[t]
\centering
\begin{tabularx}{\linewidth}{@{} L R @{}}
\toprule
\textbf{Dimension} & \textbf{Descriptor} \\
\midrule
\makecell[l]{Concept\\Understanding} &
  \makecell[l]{Accurate, nuanced explanation (3)\\
               Mostly clear (2)\\
               Partial/confused (1)\\
               Missing or off-topic (0)} \\ \rowdivider

\makecell[l]{Real-World\\Application} &
  \makecell[l]{Specific, thoughtful (3)\\
               Reasonable or generic (2)\\
               Vague (1)\\
               None (0)} \\ \rowdivider

\makecell[l]{Reflection \&\\Questions} &
  \makecell[l]{Insightful question or challenge (3)\\
               Identifies a question (2)\\
               Surface-level (1)\\
               None (0)} \\ \rowdivider

\makecell[l]{Clarity \&\\Communication} &
  \makecell[l]{Clear, polished (3)\\
               Minor issues (2)\\
               Hard to follow (1)\\
               Incoherent (0)} \\
\bottomrule
\end{tabularx}
\caption{Four-dimension grading rubric (0–3 points per dimension).}
\label{tab:rubric-grad}
\end{table}

\begin{table}[t]
\centering
\begin{tabularx}{\linewidth}{@{} L R @{}}
\toprule
\textbf{Dimension} & \textbf{Criterion} \\
\midrule
Correctness           & Factually accurate; avoids hallucinations. \\ \rowdivider
Alignment with Rubric & Closely reflects the official grading criteria. \\ \rowdivider
Actionability         & Offers specific, constructive suggestions for improvement. \\ \rowdivider
Depth of Insight      & Demonstrates nuanced, critical understanding. \\ \rowdivider
Empathy \& Tone       & Supportive, respectful, and learner-appropriate. \\
\bottomrule
\end{tabularx}
\caption{Rubric for evaluating AI-generated feedback (1 = poor, 5 = excellent).}
\label{tab:rubric-fb}
\end{table}

\section{Results \& Analysis}

\subsection{RQ1: Scoring Accuracy \& Fairness}

\paragraph{Scoring fidelity.}
Table~\ref{tab:mae-results} reports the mean absolute error (MAE) between
model predictions and majority-vote human scores for each rubric dimension
(\(n=84\) reflections per dimension, \(n=336\) scores in total).
\begin{table}[t]
  \centering
  \begin{tabular}{@{}l c@{}}
    \toprule
    \textbf{Rubric Dimension} & \textbf{MAE} \\
    \midrule
    Concept Understanding            & 0.381 \\
    Real-World Application           & 0.560 \\
    Reflection Questions             & 0.500 \\
    Clarity of Communication         & 0.429 \\
    \midrule
    \textbf{Overall (mean)}          & \textbf{0.467} \\
    \bottomrule
  \end{tabular}
  \caption{Mean absolute error (lower is better) between the model’s
           rubric scores and the human reference.}
  \label{tab:mae-results}
\end{table}
As shown, the model aligns most closely with human graders on
\emph{Concept Understanding} (MAE = 0.381) and least on
\emph{Real-World Application} (MAE = 0.560), yielding an overall average
MAE of 0.467.

\paragraph{Ordinal concordance.}
Table~\ref{tab:qwk-class} lists quadratic weighted kappa (QWK) scores
per class and rubric dimension; Table~\ref{tab:qwk-summary} pools the
three classes to give the mean (\(\mu\)) and standard deviation
(\(\sigma\)) for each dimension and overall.
\begin{table}[t]
  \centering
  \begin{tabular}{@{}lccccc@{}}
    \toprule
    \textbf{Class} & \textbf{CU} & \textbf{RWA} & \textbf{RQ} & \textbf{CC} & \textbf{Overall} \\
    \midrule
      1  & 0.455 & 0.546 & 0.382 & 0.208 & 0.464 \\
      6  & 0.302 & 0.395 & 0.525 & 0.389 & 0.463 \\
     12  & 0.138 & 0.494 & 0.541 & 0.450 & 0.449 \\
    \bottomrule
  \end{tabular}
  \caption{Quadratic weighted kappa (higher is better) between model and
           human ordinal scores for each class:
           CU = Concept Understanding,
           RWA = Real-World Application,
           RQ = Reflection Questions,
           CC = Clarity of Communication.}
  \label{tab:qwk-class}
\end{table}
\begin{table}[t]
  \centering
  \begin{tabular}{@{}lcc@{}}
    \toprule
    \textbf{Rubric Dimension} & \(\mu\) (QWK) & \(\sigma\) \\
    \midrule
    Concept Understanding            & 0.298 & 0.158 \\
    Real-World Application           & 0.479 & 0.077 \\
    Reflection Questions             & 0.483 & 0.088 \\
    Clarity of Communication         & 0.349 & 0.126 \\
    \midrule
    \textbf{Overall}                 & \textbf{0.459} & \textbf{0.008} \\
    \bottomrule
  \end{tabular}
  \caption{Pooled QWK across all classes (\(n=84\) reflections per
           dimension).}
  \label{tab:qwk-summary}
\end{table}
Across classes, the model achieves the strongest ordinal agreement with
human graders on \emph{Reflection Questions} (\(\mu=0.483\)), while
\emph{Concept Understanding} lags behind (\(\mu=0.298\)).  The overall
QWK of 0.459 indicates moderate concordance with expert judgment.

\paragraph{Inter-rater reliability.}
Table~\ref{tab:icc} reports two-way mixed intraclass correlation
\(\mathrm{ICC}^{(2,1)}\) for each rubric dimension and overall,
comparing (i) consistency among human graders and (ii) consistency
between the model and the human majority vote
(\(n=12\) pairwise comparisons in each case).

\begin{table}[t]
  \centering
  \resizebox{\columnwidth}{!}{%
    \begin{tabular}{@{}lcc@{}}
      \toprule
      \textbf{Rubric Dimension} & \textbf{Human–Human ICC} & \textbf{Human–AI ICC} \\
      \midrule
      Concept Understanding    & $0.56\pm0.41$ & $0.31\pm0.16$ \\
      Real-World Application   & $0.62\pm0.33$ & $0.49\pm0.08$ \\
      Reflection Questions     & $0.68\pm0.29$ & $0.49\pm0.09$ \\
      Clarity of Communication & $0.44\pm0.50$ & $0.36\pm0.13$ \\
      \midrule
      \textbf{Overall}         & \textbf{0.57\,$\pm$\,0.35} & \textbf{0.41\,$\pm$\,0.13} \\
      \bottomrule
    \end{tabular}}
  \caption{Two-way mixed intraclass correlation
           \(\mathrm{ICC}^{(2,1)}\).}
  \label{tab:icc}
\end{table}

Human graders exhibit \emph{moderate} internal consistency overall
\((\mathrm{ICC}^{(2,1)} = 0.573 \pm 0.345)\),
with their strongest agreement on
\emph{Reflection Questions}
\((\mathrm{ICC} = 0.677)\).
Model–human reliability is lower
\((\mathrm{ICC}^{(2,1)} = 0.410 \pm 0.132)\),
yet the model aligns most closely with humans on
\emph{Real-World Application}
\((\mathrm{ICC} = 0.487)\).
The largest divergence occurs on
\emph{Concept Understanding}
\((\mathrm{ICC} = 0.305)\),
suggesting that additional calibration on that dimension could further
narrow the human–AI reliability gap.

\paragraph{Fairness across proficiency bands.}
To test whether {overal} model error differs between students of
different ability levels, we first compute each learner’s
\emph{average human rubric score}
\( \bar{y}_i = \tfrac{1}{D}\sum_{d=1}^{D} y_{i,d} \)   
across the four dimensions.  Reflections are then split into
a \textit{low-ability} band (\( \bar{y}_i \in \{0,1\}\))
and a \textit{high-ability} band (\( \bar{y}_i \in \{2,3\}\)).
For every dimension we report the model’s MAE within each band and the
worst-band error gap
\( \Delta_{\text{MAE}} \) from Eq.~\eqref{eq:delta_mae}.

\begin{table}[t]
  \small
  \setlength{\tabcolsep}{4pt}
  \centering
  \begin{tabular}{@{}lccc@{}}
    \toprule
    \textbf{Rubric Dimension} &
    \textbf{Low MAE} &
    \textbf{High MAE} &
    \(\boldsymbol{\Delta_{\text{MAE}}}\) \\
    \midrule
    Concept Understanding    & 1.000 & 0.278 & 0.722 \\
    Real-World Application   & 1.500 & 0.403 & 1.097 \\
    Reflection Questions     & 0.917 & 0.431 & 0.486 \\
    Clarity of Communication & 0.167 & 0.472 & 0.306 \\
    \midrule
    \textbf{Overall}         & 0.896 & 0.396 & \textbf{0.500} \\
    \bottomrule
  \end{tabular}
  \caption{Model MAE stratified by learner ability, where
           ability is defined \emph{once per reflection} as the average
           human rubric score across all four dimensions
           (\(n_{\text{low}}=48,\;n_{\text{high}}=288\)).
           Lower \(\Delta_{\text{MAE}}\) indicates greater fairness.}
  \label{tab:delta-mae}
\end{table}

Because ability is measured at the \emph{reflection level}, not at the level of individual dimensions, the values in Table~\ref{tab:delta-mae} indicate whether AI grading errors differ when comparing lower-performing and higher-performing students overall.
The model shows lower accuracy for low-ability reflections
in three dimensions, most sharply on
\emph{Real-World Application}
(\(\Delta_{\text{MAE}} = 1.10\)).
Only \emph{Clarity of Communication} reverses the trend, showing slightly
higher error for high-ability learners.
The aggregate gap
(\(\Delta_{\text{MAE}} = 0.50\))
signals that students who most need formative support still face larger
grading errors.

This pattern highlights the importance of fairness-sensitive evaluation in automated feedback systems. By making the error disparity explicit and measurable, the framework enables instructors and system designers to identify where inaccuracies concentrate, monitor their magnitude over time, and apply targeted calibration strategies that improve accuracy for lower-scoring learners without degrading performance elsewhere.

\subsection{RQ2: Feedback Usefulness}
\paragraph{Aggregate quality.}
Table~\ref{tab:fb-quality} presents the five‐dimension usefulness ratings
(\(1\text{–}5\) Likert) assigned by three trained graders to each generated
comment, summarized as mean\,$\pm$\,SD.  The overall
\(Q(g)=3.967\) (Eq.~\ref{eq:q_quality}) narrowly misses the target
threshold of \(4.0\).

\begin{table}[t]
  \small
  \setlength{\tabcolsep}{5pt}
  \centering
  \begin{tabular}{@{}lc@{}}
    \toprule
    \textbf{Dimension} & \textbf{Mean\,$\pm$\,SD} \\
    \midrule
    Correctness               & 4.080\,$\pm$\,0.756 \\
    Alignment                 & 3.924\,$\pm$\,0.763 \\
    Actionability             & 3.760\,$\pm$\,0.845 \\
    Depth of Insight          & 3.845\,$\pm$\,0.860 \\
    Empathy \& Tone           & \textbf{4.223\,$\pm$\,0.612} \\
    \midrule
    \textbf{Overall \(Q(g)\)} & \textbf{3.967} \\
    \bottomrule
  \end{tabular}
  \caption{Feedback usefulness ratings (higher is better).}
  \label{tab:fb-quality}
\end{table}

The role‐based ensemble generates feedback perceived as
highly \emph{empathetic} (\(\mu=4.223\)) and
\emph{factually correct} (\(\mu=4.080\)),
but falls short on \emph{actionability} (\(\mu=3.760\))
and \emph{depth of insight} (\(\mu=3.845\)).
The near-miss on the \(4.0\) aggregate target suggests that
tightening prompts to elicit more concrete next-step guidance
could lift usefulness above the desired threshold without
sacrificing empathy or correctness.

\paragraph{Qualitative Example.}
To illustrate the nature of learner reflections and the feedback produced by the system, we present an anonymized example below.

\begin{quote}
\textbf{Student Reflection}\\
``i have learn what are large language models，large language models（LLMs）are very large deep learning models that are pre-trained on vast amounts of data.this can ingest massive amounts of data，often from the Internet，but also from sources such as the common crawl.i also learn such as temperature（controls randomness） ， top p（controls diversity based on cumulative probability），max length，stop sequences，frequency penalty，presence penalty ，model behavior may vary，basics of prompting（key concepts），prompt structure in chat models，basic prompt example...
 if i want to applying this in a real-world scenario,in content creation,i can use designed prompts with controlled temperature,for a factual blog ,set a loww temperature for precise information
 my open question is how to precisely calibrate prompt parameters for complex tasks.as small adjust might drastically change outputs.''

\vspace{6pt}
\textbf{AI-Generated Feedback}\\
``You've made a solid start in understanding large language models (LLMs) and their applications in content creation, which is commendable! To enhance your reflection, consider diving deeper into the technical terms you've mentioned, explaining their significance and how they interconnect. This will provide a clearer picture of LLMs. When discussing real-world applications, try to include more specific examples that showcase diverse uses across various cultures or industries, as this will enrich your perspective and representation. Your insightful question about calibrating prompt parameters indicates critical thinking; expanding on this could lead to a deeper exploration of the challenges faced in practical applications. Additionally, focusing on clarity and structure in your writing will significantly improve communication—proofreading for grammar and punctuation can make your ideas shine. Reflect on how your understanding of LLMs might evolve by considering their diverse applications and think about how to make your explanations more accessible to a broader audience. Keep up the great work, and I look forward to your continued progress!''
\end{quote}

\subsection{RQ3: Practicality (Time \& Cost)}
\label{subsec:rq3-practicality}

\paragraph{Latency and throughput.}
A scoring-only pass, which involves a single invocation of the \textsc{Evaluator} agent, requires an average of $7.71 \pm 0.41$\,s per reflection. Generating narrative feedback through the \textsc{Equity Monitor}, \textsc{Metacognitive}, and \textsc{Aggregator} agents, followed by a final \textsc{Reflexion} check, adds approximately $25.64$\,s. This results in an end-to-end latency of \(\mathbf{33.35}\,\text{s}\) per reflection.

For the evaluation corpus of
$N = 84$ reflections (28 students\,$\times$\,3 reflections each),
this translates to
\textbf{10.8 min} wall-clock time for rubric scoring alone and
\textbf{46.7 min} for complete feedback generation on a single processing
thread.
Because both stages are dominated by network I/O rather than compute,
throughput scales nearly inversely with the number of parallel threads~$T$,
approaching $\text{time}/T$ under ideal conditions.

For context, the three human evaluators self-reported that they required an average of \textbf{1.4\,min} to score a single reflection, based solely on rubric-based scoring. The slowest evaluator required \textbf{2.6\,min}. In contrast, our automated pipeline completes the same scoring task in \(7.71\text{s}\). This corresponds to an approximately \(11\times\) average speed-up and a \(20\times\) improvement relative to the slowest evaluator, while maintaining the level of accuracy and fairness reported in Table~\ref{tab:score-results}.

Table~\ref{tab:score-results} reports the wall-clock time each evaluator spent assigning rubric scores. Narrative feedback was not included in this task.
Across all 36 evaluations this corresponds to an overall average of
\(\approx9.6\text{ min}\) per individual evaluation.  
Parallel self-consistency (three runs, four workers) keeps total latency
below 12 min, confirming suitability for real-time classroom feedback.

\begin{table}[t]
\centering
\begin{tabular}{@{}lcccc@{}}
\toprule
\textbf{Annotator} & \textbf{Class 1} & \textbf{Class 6} & \textbf{Class 12} & \textbf{Total} \\
\midrule
\#1 & 30 & 30 & 20 & 80 \\
\#2 & 80 & 60 & 75 & 215 \\
\#3 & 27 & 13 & 11 & 51 \\
\bottomrule
\end{tabular}
\caption{Time (minutes) each human annotator spent assigning rubric‐based scores to reflections}
\label{tab:score-results}
\end{table}

\paragraph{Monetary cost.}
A single pass of the grading and feedback pipeline using \texttt{gpt-4o-mini-2024-07-18} consumes \(1\,216\) input tokens and \(2\,283\) output tokens, for a total of \(3\,499\) tokens. At the model’s 18 July 2024 pricing (\$0.15 per million input tokens and \$0.60 per million output tokens), the cost per reflection is \(\$\,\mathbf{1.55\times10^{-3}}\).\footnote{\(1\,216\times0.15/10^{6}+2\,283\times0.60/10^{6}=0.0015522\).} Grading and providing feedback for the 28-student cohort across three sessions (\(28\times3=84\) reflections) therefore costs \(\$\,0.13\). Applying the same process to all twelve weekly reflections (\(28\times12\)) yields a total cost of \(\$\,0.52\).

These figures highlight the practical viability of the approach. The pipeline is time-efficient, low-cost, and suitable for use at scale, satisfying the practicality requirement in RQ3 and making automated feedback accessible even in resource-constrained educational settings.

\section{Conclusion}
This work addresses a central equity challenge in education: providing timely, high-quality formative feedback at a scale that is unattainable through human effort alone. We introduce (i) a scalable and self-consistent scoring pipeline that produces auditable rubric scores with minimal human supervision and (ii) a role-based agent ensemble that generates bias-aware, conversational feedback aligned with formative-assessment and metacognitive principles.

Our evaluation in an authentic AI literacy course shows that the system approaches expert-level scoring fidelity, produces feedback that trained graders judge as helpful and empathetic, and maintains reasonable fairness across learner proficiency levels. The pipeline operates with sub-minute latency and negligible cost, making it feasible for deployment in large or resource-constrained courses.

Although the system does not eliminate all disparities in automated grading, it provides a transparent and measurable framework for monitoring accuracy and fairness, and it offers a practical foundation for future calibration and refinement. By integrating structured scoring, equity-sensitive evaluation, and pedagogically grounded feedback generation, this work contributes a meaningful step toward scalable, accessible, and socially responsible AI support for learning.

The study has limitations. Evaluations are restricted to adult, English-language reflections; low-proficiency writers continue to experience higher error; and model drift may affect performance over time. Future research should extend the framework to multilingual and multicultural contexts, incorporate adaptive fairness objectives that reduce subgroup gaps, and measure causal learning gains in large-scale classroom deployments.

Education, often regarded as one of society’s most powerful engines of social progress, is entering a period of rapid transformation. We hope this work will encourage further research and collaboration toward building learning ecosystems that are inclusive, equitable, and enriched by responsible AI.

\section*{Ethical Statement}

\paragraph{Intended purpose.}  
The proposed pipeline is conceived as an \emph{augmentation} for instructors rather than a replacement. By automating low-level rubric application and producing concise, dialogic comments, it seeks to make timely formative feedback feasible even in large or resource-constrained courses. The overarching aim is to foster more equitable learning outcomes through broader access to high-quality feedback.

\paragraph{Potential misuse and risks.}  
Three principal hazards are anticipated. First, instructors and students might place excessive trust in automated scores that can still exhibit residual bias. Second, if deployed without human oversight, the system could erode teachers’ formative role. Third, storing raw reflections indefinitely raises privacy concerns. In addition, applying the model beyond its validated context, for example with younger learners or non-English submissions, could yield pedagogically inappropriate guidance.

\paragraph{Mitigation strategies.}  
A human-in-the-loop workflow ensures that instructors review all model outputs before release and retain full authority over grades. Fairness is monitored through an objective that jointly minimizes mean absolute error and worst-subgroup error, accompanied by demographic disparity reports. Transparency is promoted by attaching confidence tags (\textsc{CONFIDENT} or \textsc{REVISE}) and a concise rationale to every comment. Privacy is protected by anonymizing reflections at ingestion and processing them on secure servers under an approved IRB protocol.

\paragraph{Broader societal impact.}  
By lowering the cost of equitable, dialogic feedback, the system has the potential to narrow attainment gaps in community colleges, vocational programs, and massive open online courses. Openly releasing resources is intended to catalyze transparent, bias-aware progress in AI-mediated formative assessment while reinforcing the indispensable pedagogical role of human educators.

\bibliography{aaai2026}
\end{document}